\documentclass[prd,showpacs,preprintnumbers,amsmath,amssymb]{revtex4}

\usepackage{graphicx}
\usepackage{dcolumn}
\usepackage{bm}

\def\be{\begin{eqnarray}}
\def\ee{\end{eqnarray}}
\def\bea{\begin{eqnarray}}
\def\eea{\end{eqnarray}}

\def\pp{{\bf p}_\perp}

\def\xT{{\bf x}_\perp}

\def\bT{{\bf b}_\perp}
\def\bp{{\bf b}_\perp}

\def\RT{{\bf R}_\perp}

\def\0T{{\bf 0}_\perp}
\begin{document}


\title{Anomalous Magnetic Moments and Quark Orbital Angular Momentum}

\author{Matthias Burkardt}
 \affiliation{Department of Physics, New Mexico State University,
Las Cruces, NM 88003-0001, U.S.A.}
\author{Gunar Schnell}

\affiliation{Subatomaire en Stralingsfysica, Universiteit Gent, 9000 Gent,
  Belgium} 
\date{\today} 

\begin{abstract}
We derive an inequality for the distribution of quarks 
with nonzero orbital angular momentum, and thus demonstrate, in a 
model-independent way, that a nonvanishing anomalous magnetic moment
requires both a nonzero size of the target as well as the presence
of wave function components with quark orbital angular momentum $L^q_z>0$.
\end{abstract}

\maketitle

\section{Introduction}
Detailed measurements of the spin structure of the nucleon
\cite{EMC,SMC,SMC:deuteron,SMC:p+d,SMC:lowX,COMPASS,E143,E155:d,E155:Q2evolution,hermes:g1p,jlab:hallA}
have revealed that only a small fraction of the 
nucleon spin is carried by the quark spin. This result immediately
raised the question, which degrees of freedom carry the rest.
Unfortunately, both orbital angular momentum and the gluon spin
are difficult to access experimentally, and therefore little
rigorous information exists about quark orbital angular momentum.

Meanwhile, many qualitative statements regarding orbital angular
momentum have been made. For example, when one expresses the
matrix element for the anomalous magnetic moment of the nucleon
in terms of light-cone wave functions (summed over all Fock 
components), a nonzero anomalous magnetic moment can only result
when there is a nonzero probability that the vector current flips
the nucleon helicity \cite{Brodsky:1980zm,ELz}. 
Since the same matrix element conserves
the spin of the quarks it is evident that some orbital angular
momentum must be transfered to the quarks. Hence a nonzero
anomalous magnetic moment can only occur when the target wave function
contains components with nonzero orbital angular momentum.
While this argument is rigorous, it leaves open quantitative questions
regarding the norm of those wave function components or
perhaps the resulting net $L_z^q$.
Within models, it has also been found that a point-like object cannot
produce a nonzero anomalous magnetic moment \cite{Brodsky:1980zm,schlumpf} 
and within this model one can even derive quantitative bounds.

Similarly, the observation of a nonzero Sivers \cite{sivers}
effect by the HERMES collaboration \cite{hermes} seems to indicate
wave-function components with nonzero $L_z^q$ since the effect
requires an interference between initial nucleon states that have
opposite helicity. Furthermore, orbital angular momentum seems to
play a central role in all models for the Sivers function 
\cite{Boros:1993ps,hwang,model,Yuan:2003wk,Bacchetta:2003rz}.

The main purpose of this note is to make some of the statements
regarding the anomalous magnetic moment more 
quantitative. In order to accomplish this goal, we start from
the matrix element that yields the generalized parton distribution 
$E^q$ and apply the Cauchy-Schwarz inequality which will then provide
a lower bound on the norm of wave function components with nonzero 
orbital angular momentum.

\section{Decomposition of the Nucleon Spin}
In QCD, there is no unique way to decompose the nucleon spin into
quark spin, quark angular momentum, gluon spin, and gluon orbital
angular momentum. For example, Ji has considered a decomposition
based on the $M^{012}$ component of the angular momentum
tensor \cite{JiPRL}
\be
M^{012}= \frac{i}{2}q^\dagger \left({\vec r} \times {\vec D}
\right)^zq
+\frac{1}{2}q^\dagger \sigma^zq + 2 \mbox{Tr} E^j
\left({\vec r} \times {\vec D} \right)^z A^j + \mbox{Tr}
\left({\vec E} \times {\vec A} \right)^z .
\label{M012}
\ee
The matrix elements of the four terms in Eq.~(\ref{M012}) are
interpreted as the quark orbital angular momentum, quark spin, 
gluon orbital angular momentum, and gluon spin respectively.
One advantage of this decomposition is gauge invariance, another is
the fact that the matrix elements of the first term in Eq.~(\ref{M012}) 
can be probed in deeply virtual Compton scattering.
A disadvantage of this decomposition is that the orbital angular
momenta in Eq.~(\ref{M012}) contain interactions through the gauge
covariant derivative.

An alternative decomposition of the total angular momentum starts
from $M^{+12}$ in light cone gauge $A^+\equiv A^0+A^z=0$~\cite{JaffeLz}
\be
M^{+12} = \frac{1}{2}q^\dagger_+\left({\vec r}\times i{\vec \partial}
\right)^z q_+ + \frac{1}{2}q^\dagger_+ \gamma_5 q_+
+ 2 \mbox{Tr} F^{+j}\left({\vec r}\times i{\vec \partial} 
\right)^z A^j
+ \varepsilon^{+-ij}\mbox{Tr}F^{+i}A^j
\label{M+12}
\ee
Here $q_+ \equiv \frac{1}{2} \gamma^-\gamma^+ q$ is the dynamical
component of the quark field operators. The obvious disadvantage
of using Eq.~(\ref{M+12}) to interpret the total angular momentum of 
the nucleon, which we all know is equal to $\frac{1}{2}\hbar$, as a 
sum of quark and gluon orbital and spin angular momentum,
respectively, is the fact that Eq.~(\ref{M+12}) is not gauge 
invariant. However, it is invariant under the residual group
of gauge transformations that leave $A^+=0$ and where the $A^j$
satisfy antisymmetric boundary conditions at $x^-=\pm \infty$.
The main advantages of using Eq.~(\ref{M+12}) to decompose the 
nucleon spin is the fact that all terms in Eq.~(\ref{M+12}) are
quadratic in the fields. Unlike Eq.~(\ref{M012}) there is therefore
no ambiguity as to the interpretation of interaction terms.
Another advantage is the fact that the matrix elements of 
the various terms in Eq.~(\ref{M+12}) can be easily expressed in
terms of light-cone wave functions. 

We should emphasize that although we have listed here some of the 
most obvious advantages and disadvantages of these two possibilities
for decomposing the angular momentum, we do not consider one or
the other superior in general. However, for specific applications
one of these two decompositions may be advantageous. In particular,
if one wants to place constraints on light-cone wave functions,
which find many applications in hadron phenomenology, then one may
prefer the decomposition of the orbital angular momentum based
on $M^{+12}$ (\ref{M+12}). In the rest of this paper we will
exclusively study the light-cone decomposition of $J_z$, based on
Eq.~(\ref{M+12}).

We thus consider in the following
the orbital angular momentum of quarks with flavor
$q$
\be
L_z^q = \int\! dx^-\!\!\int d^2\xT
\frac{1}{2}q^\dagger_+\left({\vec r}\times i{\vec \partial}
\right)^z q_+ 
\label{Lz}
\ee 
and we will investigate its role in nucleon spin-flip matrix
elements, such as the anomalous magnetic moment.

As it stands, Eq.~(\ref{Lz}) is gauge dependent. We therefore
need to specify the gauge. First of all, we impose the light-cone 
gauge condition $A^+=0$. However, this does not yet completely
fix the gauge since $A^+$ remains zero under gauge transformations
with a phase that only depends on $\xT$. We thus impose as an additional
condition that the $\perp$ component of the gauge field satisfies 
anti-periodic boundary conditions at light-cone infinity
\be
{\bf A}_\perp(x^-=\infty,\xT) = - {\bf A}_\perp(x^-=-\infty,\xT) 
\label{bc}.
\ee
With this additional condition, the definition~\eqref{Lz} of $L_z^q$ 
becomes unique.

Although all results that we will be deriving will be valid in this
gauge only, we believe they will nevertheless be useful since
a lot of hadron phenomenology is based on light-cone wave functions
and their use usually implies the use of light-cone gauge.
It is therefore fair to say that although our results will not have a
gauge-independent interpretation, they will nevertheless provide
additional insights about light-cone wave functions of hadrons.

\section{Anomalous Magnetic Moment and GPDs}

We first consider the generalized parton distribution 
$E(x,0,-{\bf \Delta}_\perp^2)$, which appears in non-forward
nucleon
spin-flip matrix elements of light-cone correlation functions. For
purely transverse momentum transfer $p^+={p^\prime}^+$ one finds
\be
\label{defH}
\left\langle P\!\!+\!\!\Delta,\! 
\uparrow\!\left|O^q(x,{\bf 0}_\perp) 
\right| P,\!\uparrow\right\rangle
&=&H^q(x,\!0,\!-{\bf \Delta}_\perp^2)\\
\label{defE}
\left\langle P\!\!+\!\!\Delta,\! 
\uparrow\!\left|O^q(x,{\bf 0}_\perp) 
\right| P,\!\downarrow\right\rangle
&=& 
-\frac{\Delta_x\!\!-\!i\Delta_y}{2M}E^q(x,\!0,\!-{\bf \Delta}_\perp^2)\ee
where
\be 
\label{Oq}
O^q(x,\bp)=
\int \frac{dx^-}{4\pi} e^{ip^+x^- x}
\bar{q}\!\left(0^-,\bp \right)\gamma^+
q\!\left({x^-},\bp \right)
\ee
(in gauges other than light-cone gauge $A^+=0$ one
needs to insert a Wilson line gauge link in Eq.~(\ref{Oq})).
In the forward limit, integration over $x$ yields the anomalous
magnetic moment contribution from quarks with flavor $q$
\be
\int\! dx E^q(x,0,0)= \kappa^q .
\ee
Since the operator in Eq.~(\ref{defE}) is chirally even, the matrix
element is diagonal in quark spin. Since the matrix element involves
a nucleon spin-flip, angular momentum conservation thus requires
a change in quark orbital angular momentum.
As a result, the mere fact that the anomalous magnetic moment
of the nucleon is nonzero implies that there must be components
in the nucleon wave function that have a nonzero orbital angular
momentum \cite{ELz} (of course, this does not necessarily mean 
that there is any {\sl net} orbital angular momentum).
In the following we will try to make this statement more 
quantitative, i.e., we will attempt to place constraints on the
probability to find nonzero orbital angular momentum components
in the nucleon wave function.

In order to facilitate the separation between intrinsic orbital
angular momentum and the orbital angular momentum due to the
motion of the entire nucleon, we first switch to a representation
of states that are eigenstates of the transverse center of momentum
\be
\left| p^+,{\bf R}_\perp,\lambda\right\rangle \equiv {\cal N}
\int d^2\pp \left| p^+,{\bf p}_\perp,\lambda\right\rangle
e^{i\bT \cdot \RT},
\ee
where ${\cal N}$ is a normalization constant.
In this basis, we can define impact parameter dependent parton
distributions as
\be
{\cal H}^q(x,\bT) &=& \left\langle p^+,{\bf R}_\perp,\uparrow
\right| O^q(x,\bT)
\left| p^+,{\bf R}_\perp,\uparrow
\right\rangle
\\
\left(\frac{\partial}{\partial b_x} -i
\frac{\partial}{\partial b_y}\right) 
\frac{{\cal E}^q(x,\bT)}{2M} &=& \left\langle p^+,{\bf R}_\perp,
\uparrow
\right| O^q(x,\bT)
\left| p^+,{\bf R}_\perp,\downarrow\right\rangle .
\ee
The impact parameter dependent PDFs are related to GPDs via a simple
Fourier transform \cite{koenig,Burkardt:2002hr,Diehl:2002he,Belitsky:2003nz} 
\be
{\cal H}(x,\bT) &=& \int \frac{d^2{\bf \Delta}_\perp}{(2\pi)^2} e^{i\bT \cdot{\bf \Delta}_\perp}
H(x,0,-{\bf \Delta}_\perp^2)\\
{\cal E}(x,\bT) &=& \int \frac{d^2{\bf \Delta}_\perp}{(2\pi)^2} e^{i\bT \cdot{\bf \Delta}_\perp}
E(x,0,-{\bf \Delta}_\perp^2) ,
\ee
and the normalization is such that $\int d^2\bT q(x,\bT)=q(x)$.

Since the GPDs provide simultaneous information
about the longitudinal momentum and the transverse position of partons,
it is instructive to introduce creation operators in this ``hybrid
space''. For example, for $x>0$ we define destruction operators
for quarks with momentum fraction $x$ at transverse position $\bT$
\be
b(xp^+,\bT) = \int dx^- e^{ip+x^-x} q_{+}(x^-,\bT)
\ee
and similarly for anti-particles ($x<0$). 
For now we will suppress the
helicity of the quarks and consider only quantities 
that have been summed over the 
helicities of the quarks. In terms of these, the
impact parameter dependent PDFs take on a particularly simple
form ($x>0$)
\be
{\cal H}(x,\bT) &=& 
\left\langle p^+,{\bf R}_\perp,\uparrow
\right| b^\dagger(xp^+,\bT)b(xp^+,\bT)
\left| p^+,{\bf R}_\perp,\uparrow
\right\rangle\\
\frac{1}{2M}\left(\frac{\partial}{\partial b_x} +i 
\frac{\partial}{\partial b_y}\right)
{\cal E}(x,\bT) &=& 
\left\langle p^+,{\bf R}_\perp,\uparrow
\right| b^\dagger(xp^+,\bT)b(xp^+,\bT)
\left| p^+,{\bf R}_\perp,\downarrow
\right\rangle,
\label{Eimpact}
\ee
which emphasizes their physical interpretation as densities.
Upon introducing
\be
B_+^q(x)\equiv \int d^2\bT \left(b_x+ib_y\right)
b^\dagger(xp^+,\bT)b(xp^+,\bT)
\ee
and integration by parts we find
\be
\left\langle p^+,{\bf R}_\perp,\uparrow
\right| B_+^q(x)
\left| p^+,{\bf R}_\perp,\downarrow \right\rangle
 = \frac{1}{2M} E^q(x,0,0).
\label{E}
\ee
We can use this result to provide a
formal proof that $\Delta L_z^q=1$ in the matrix element 
defining $E^q(x,0,0)$. Indeed, one easily verifies the commutation 
relation
\be
\left[L_z^q,B_+^q(x)\right]=B_+^q(x),
\label{raise}
\ee
which proves that $B_+^q$ has nonvanishing matrix elements
only between states
that differ by one unit of orbital angular
momentum $L_z^q$ for flavor $q$. This observation is consistent with
results based on overlap integrals of light-cone wave functions \cite{ELz}.

\section{Angular momentum decomposition of PDFs}
In order to derive some quantitative 
constraints on the light-cone wave functions of hadrons, we first introduce an
angular momentum  
decomposition for parton distributions, i.e.,
\be
b^\dagger(xp^+,\bp) = \sum_m b_m^\dagger(xp^+,\bp)
\ee
where
\be
b_m^\dagger(xp^+,\bp) =  e^{+im\phi_b}
\int_0^{2\pi} \frac{d\phi_b^\prime}{2\pi}
e^{-im\phi_b^\prime}b^\dagger(xp^+,\bp^\prime)
\label{creator}
\ee
where $b_x= |\bp| \cos \phi$ and $b_y= |\bp| \sin \phi$, and 
$b_x^\prime = |\bp| \cos \phi_b^\prime$ and $b_y^\prime= 
|\bp| \sin \phi_b^\prime$, respectively. 

Of course, the above decomposition into orbital angular momentum components 
is scale dependent due to the scale dependence of light-cone wave 
functions \cite{Jiscale}. However, we imagine performing such a decomposition
at a fixed scale at which we perform the analysis of this decomposition.

The physical
interpretation of the creation and destruction operators (\ref{creator})
is that they
create quarks with $m$ units of angular momentum (in the z-direction),
i.e.,
\be
\left[L_z^q, b_m^\dagger(xp^+,\bp)\right] = m b_m^\dagger(xp^+,\bp).
\ee
In terms of these angular momentum projected creation operators,
the operator $B^q_+$ appearing in the impact parameter space
representation of the matrix element for the
anomalous magnetic moment takes on the form
\be
B^q_+ = \sum_m \int d^2\bT \left(b_x+ib_y\right)
b_{m+1}^\dagger(xp^+,\bT)b_m(xp^+,\bT).
\label{mexpansion}
\ee
Inserting the angular mode expansion into Eq.~(\ref{E}) and applying
the Cauchy-Schwarz inequality to scalar products between states
yields
\be
\frac{ E^q(x,0,0)}{2M}&=&
\sum_m \int d^2\bT \left(b_x+ib_y\right)
\left\langle p^+,{\bf 0}_\perp,\uparrow
\right|
b_{m+1}^\dagger(xp^+,\bT)b_m(xp^+,\bT)
\left| p^+,{\bf 0}_\perp,\downarrow \right\rangle \nonumber\\
&\leq& 
\sum_m \int d^2\bT \left|\bT\right|
\sqrt{  \left\langle p^+,{\bf 0}_\perp,\uparrow
\right|b_{m+1}^\dagger(xp^+,\bT)b_{m+1}(xp^+,\bT)
\left| p^+,{\bf 0}_\perp,\uparrow \right\rangle}
\nonumber\\
& &\quad \quad \quad \quad \quad \quad \quad \times\sqrt{
\left\langle p^+,{\bf 0}_\perp,\downarrow\right|
b_{m}^\dagger(xp^+,\bT)b_m(xp^+,\bT)
\left| p^+,{\bf 0}_\perp,\downarrow \right\rangle}
.
\label{CSU1}
\ee
In order to simplify the notation, we now introduce
the distribution of partons with orbital angular momentum $m$
in a target with spin $\uparrow$
\be
q^\uparrow_m(x) \equiv \int d^2\bT\left\langle p^+,{\bf 0}_\perp,\uparrow
\right|b_{m}^\dagger(xp^+,\bT)b_{m}(xp^+,\bT)
\left| p^+,{\bf 0}_\perp,\uparrow \right\rangle
\ee
as well as the ${\bf b}_\perp^2$-weighted 
distribution of partons with orbital angular momentum $m$
\be
{b^2}^\uparrow_m(x) \equiv \int d^2\bT\left\langle p^+,{\bf 0}_\perp,\uparrow
\right|b_{m}^\dagger(xp^+,\bT)b_{m}(xp^+,\bT)
\left| p^+,{\bf 0}_\perp,\uparrow \right\rangle {\bf b}_\perp^2.
\ee
Obviously we have
\be
q_m^\downarrow (x) &=& q_{-m}^\uparrow (x)\\
b^{2,\downarrow}_m(x)&=&  b^{2,\uparrow}_{-m}(x).
\ee
To each term in the sum in Eq.~(\ref{CSU1}) we now apply the 
Cauchy-Schwarz inequality for integrals
\be
\int d^2\bT \sqrt{f(\bT)g(\bT)}
\leq \sqrt{ \int d^2\bT f(\bT)}  \sqrt{ \int d^2\bT g(\bT)}.
\ee
For $m\geq 0$ we apply this inequality with
\be
f(\bT) &=& \int d^2\bT\left\langle p^+,{\bf 0}_\perp,\uparrow
\right|b_{m+1}^\dagger(xp^+,\bT)b_{m+1}(xp^+,\bT)
\left| p^+,{\bf 0}_\perp,\uparrow \right\rangle =
q_{m+1}^\uparrow (x)\\
\nonumber g(\bT) &=& \int d^2\bT\,{\bf b}_\perp^2 
\left\langle p^+,{\bf 0}_\perp,\downarrow
\right|b_{m}^\dagger(xp^+,\bT)b_{m}(xp^+,\bT)
\left| p^+,{\bf 0}_\perp,\downarrow \right\rangle =
b_m^{2,\downarrow} (x),\ee 
while for $m<0$ we identify
\be
f(\bT) &=& 
\int d^2\bT\,{\bf b}_\perp^2\left\langle p^+,{\bf 0}_\perp,\uparrow
\right|b_{m+1}^\dagger(xp^+,\bT)b_{m+1}(xp^+,\bT)
\left| p^+,{\bf 0}_\perp,\uparrow \right\rangle =
b_{m+1}^{2\uparrow}(x)\\
\nonumber g(\bT) &=& \int d^2\bT
\left\langle p^+,{\bf 0}_\perp,\downarrow
\right|b_{m}^\dagger(xp^+,\bT)b_{m}(xp^+,\bT)
\left| p^+,{\bf 0}_\perp,\downarrow \right\rangle =
q_m^{\downarrow} (x),\ee
yielding
\be 
\frac{ E^q(x,0,0)}{2M}&\leq& \sum_{m\geq 0}
\sqrt{q_{m+1}^\uparrow (x) b_m^{2,\downarrow}(x)}
+ \sum_{m<0}\sqrt{b_{m+1}^{2,\uparrow} (x) q_m^{\downarrow}(x)}
\\
&=& \sum_{m\geq 0}
\sqrt{q_{m+1}^\uparrow (x) b_{-m}^{2\uparrow}(x)}
+ \sum_{m< 0}\sqrt{b_{m+1}^{2\uparrow} (x) q_{-m}^{\uparrow}(x)}
\nonumber\\
&=& \sum_{m\geq 0}
\sqrt{q_{m+1}^\uparrow (x) b_{-m}^{2,\uparrow}(x)}
+ \sum_{m>0} \sqrt{q_{m}^{\uparrow}(x)b_{1-m}^{2,\uparrow} (x)}
\nonumber\\
&=& 2 \sum_{m\geq 0}
\sqrt{q_{m+1}^\uparrow (x) b_{-m}^{2,\uparrow}(x)}\nonumber\\
&\leq& 2 \sqrt{ \left(\sum_{m\geq 0} q_{m+1}^\uparrow (x)\right) 
\left(\sum_{n\geq 0} b_{-n}^{2,\uparrow}(x)\right)},
\ee
where in the last step we used ($f_m,g_m \geq 0$)
\be
\sum_m \sqrt{f_m g_m} \leq \sqrt{ \left( \sum_m f_m\right)
\left(\sum_ng_n \right)}.
\ee
If we now introduce the distribution of quarks with positive angular
momenta in a nucleon with spin in the $+\hat{z}$ direction
\be
q_{L_z\geq 1}(x) = \sum_{m\geq 1} q^\uparrow_m(x)
\ee
as well as the contribution to ${\bf b}_\perp^2(x)$ from
quarks with zero or negative $L_z^q$
\be
b^2_{L_z\leq 0}(x) = \sum_{m\leq 0} b^{2,\uparrow}_m(x)
\ee
our result can be cast into the form
\be
\left(\frac{ E^q(x,0,0)}{4M}\right)^2 &\leq& q_{L_z\geq 1}(x) \,
b^2_{L_z\leq 0}(x),
\label{master}
\ee
which is the main result of this paper. A slightly stronger inequality
can be obtained by keeping track of the quark helicity. For this 
purpose we note that both quark helicities contribute equally in
Eq. (\ref{E}). Upon repeating the above analysis starting from
quarks with helicity $\uparrow$, one thus arrives at
\be
\frac{ E^q(x,0,0)}{4M} &\leq& 
\sqrt{ q^\uparrow _{L_z\geq 1, \uparrow }(x) \, 
b^{2,\uparrow}_{L_z\leq 0, \downarrow}(x)}
+ \sqrt{
q^\uparrow_{L_z\geq 1, \downarrow}(x) \, b^{2,\uparrow}_{L_z\leq 0, 
\uparrow }(x)},
\label{master2}
\ee
where for example $q^\uparrow_{L_z\geq 1, \downarrow}(x)$ denotes the
distribution of quarks with helicity $\downarrow$ and positive orbital angular
momentum in a nucleon with helicity $\uparrow$. 
An even stronger version of Eq.~\eqref{master2} can be achieved following
Ref.~\cite{burkardt:PLB582} by starting with
\(\gamma^+(1\pm\gamma_5)/2\) instead of only \(\gamma^+\) in
Eq.~\eqref{Oq}. In that case only one of the two terms on the
r.h.s.~of~\eqref{master2}  remains,
and \(\frac{E}{4M}\) gets replaced by \(\frac{E}{8M}\)~\cite{diehl:private}. 
Additional simplified
inequalities can be derived if one neglects quarks with orbital angular
momenta  $|L_z^q| \geq 2$.
\section{Discussion}
First of all, our result illustrates that a nonvanishing anomalous 
magnetic moment (and hence nonvanishing $E(x,0,0)$)
implies both 
\begin{itemize}
\item wave function components with positive quark orbital
angular momentum (for $S_z = + \frac{1}{2}$)
\item a nonvanishing $\perp$ size
\end{itemize}
The fact that the size of hadrons needs to be nonzero and that nonzero
angular momentum components need to be present if a state has an
anomalous magnetic moment has been observed before in light-cone
wave function models of hadrons \cite{schlumpf,ELz}. What is new
in our paper is the fact that our result provides a 
model independent, quantitative lower bound for the $\perp$ size
distribution, which unlike in Ref.~\cite{burkardt:PLB582} is connected to the
quark orbital angular momentum. Furthermore, we have shown that there must be
quarks 
with orbital angular momentum in the same direction as the nucleon 
spin --- regardless of the sign of the anomalous magnetic moment.
This is at first surprising, since one might have expected that
if the anomalous magnetic moment is due to an orbital angular 
momentum then, when $E(x,0,0) <0$  the quark should orbit 
opposite to the nucleon spin. However, one needs to keep in mind that
we make no statement about the {\sl net} orbital angular momentum.
We only derived a lower bound on contributions to the net $L_z^q$ from modes
with $L_z^q>0$. 

The necessity of quark orbital angular momentum for an anomalous 
magnetic moment is particularly surprising from the nonrelativistic quark
model (NRQM) point of view. In the NRQM the anomalous magnetic 
moment is entirely due to the spins of the quarks and no orbital 
angular momentum is needed. However, in this
point the NRQM is not completely consistent: For example, when a 
massive Dirac particle is in some bound state then, due to the
localization, the particle must necessarily possess a nonzero momentum
and thus a nonvanishing lower Dirac component. The lower component
for a massive s-wave quark has orbital angular momentum $L_z^q=1$.
Since even the $d$ quark appears in the NRQM wave functions with
both spin up and down, this relativistic effect gives rise to
wave function components with both positive and negative
orbital angular momentum. In our inequality~(\ref{master}),
the lower bound on orbital angular momentum is proportional to
the inverse radius squared of the quark distribution. 
Strictly speaking the non-relativistic limit requires \(Rm_q
\rightarrow\infty\) and hence \(RM\rightarrow\infty\). 
In this limit the lower bound on \(q_{L_z\ge 1}\) goes to zero.  
Therefore, in a system where the nonrelativistic approximation would be
allowed, 
a nonzero anomalous magnetic moment would not necessarily require nonzero
orbital angular momentum.

What is also surprising is that, once $E(x,0,0)$ is nonzero, there need to be
quarks with positive orbital angular momentum --- regardless of
the sign of $E(x,0,0)$. This is because for $E(x,0,0)$ to be nonzero
there ought to be a nonzero matrix element between a nucleon with spin up
and a nucleon with spin down in Eq.~(\ref{E}). For example, if the active quark
in the initial state had $L_z^q=0$, then it needs to have 
$L_z^q=-1$ in the final state, i.e., $L_z^q$ in the same direction as the
nucleon 
spin. Likewise if $L_z^q=0$ in the final state, $L_z^q$ must have been $+1$ in
the 
initial state in Eq.~\eqref{E}. Similarly one can convince oneself in the
general case   
(initial and final state have nonzero $L_z^q$) that a nonzero $E(x,0,0)$ always
requires wave function components with $L_z^q>1$ in a nucleon that has
$S_z=\uparrow$. 
This is surprising since in Ji's relation the sign of $E$ seems to suggest the
sign of $L_z^q$. However, one needs to keep in mind that Ji's relation deals
with the 
{\it net} $L_z^q$ and secondly the angular momentum decomposition to which the
Ji relation applies~(\ref{M012}) 
does not have to be the same as the angular momentum decomposition
in the light-cone framework~(\ref{M+12}).

In order to get some quantitative feeling for the lower bound, we 
consider a model for \(u\)-quarks in the valence region where ($\kappa=2$)
\be
q(x)&=&8(1-x)^3\\
E(x,0,0)&=&6\kappa (1-x)^5 = 12 (1-x)^5 \\ 
M^2b^2(x)&=& M^2R^2 8(1-x)^5
\ee
where we pick $R=0.5 fm$ for the transverse size, i.e., $MR=2.5$ and we let $b^2(x)\propto
(1-x)^2 q(x)$, consistent with a finite size for large $x$.
Since no data is
available on $E(x,0,0)$ and $b^2(x)$, we simply make
an educated guess regarding these functions in order to be able to make an
order of magnitude estimate for the probability
to find nonzero orbital angular momentum. The ansatz for $E(x,0,0)$ is
motivated by the constraint that $E(x,0,0)$ must vanish by
two powers of $(1-x)$ faster than $q(x)$ for large $x$. The ansatz for $b^2(x)$  was
motivated by the constraint that  the nucleon has a
finite transverse size for large $x$.  The rest of the above ansatz was guided
by the {\it ad hoc} requirement to make the model as simple as possible and therefore
the idea is that this model should merely serve
as a guide for what order of magnitude one should expect
(``back of an envelope estimate'') with parameters that
are within the range of what is expected in the valence region.
With the above model parameters we find for the distribution of quarks
with positive orbital angular momentum
\be
\frac{9}{50}(1-x)^5 \leq q_{L_z\geq 1}(x),
\ee
i.e., only some \% compared to the distribution summed over all
orbital angular momenta. 

While this is much smaller than current estimates based
on Ji's relation (see for example Ref.~\cite{estimate} for an up to date
estimate), 
one should keep in mind that the orbital angular momentum obtained from
the Ji relation differs from the orbital angular momentum from the
light-cone decomposition (\ref{M+12}).

Although our lower bound is not very spectacular, it still provides
the first model-independent lower bound on the distribution of
quarks with positive angular momentum distribution. One may also 
wonder how such a low bound is consistent with transverse flavor
dipole moments $|d^q_y|\approx 0.2 fm$ for a moving nucleon that are 
quite significant and almost the same order of magnitude as the 
$\perp$ size of the nucleon. 
The important point here is that the $\perp$ distortion, 
which is described by our starting equation $E$, contains
a piece involving the overlap between the $L_z^q=0$ component of
the wave function and the $L_z^q=1$ component, i.e., it is
linear in the amplitude for finding a quark with nonzero orbital
angular momentum, whereas the probability for $L_z^q>0$ is quadratic
in that amplitude. The situation is thus somewhat analogous to the
quadrupole moment of the deuteron, which is linear in the d-wave
component and quite large, while the actual d-wave probability is
tiny.

\section{Summary}
We have derived an inequality, which provides a model-independent
lower bound on the norm of wave-function components with
nonzero quark orbital angular momentum. Although the numerical bound thus
obtained 
is not very strong --- requiring only a few \% probability
for nonzero orbital angular momentum -- our
result represents the first quantitative estimate for
wave-function components with nonzero $L_z^q$.
Moreover, the resulting constraints on the light-cone wave functions of the
nucleon, which only enter the distributions quadratically, are much stronger. 
The bound that we derive involves the transverse size (rms-radius)
times the mass
of the nucleon. This quantifies the known result that a
point-like particle cannot have a nonzero anomalous magnetic moment.
Unlike $L_z^q$ itself, which depends quadratically on wave-function components
with nonzero $L_z^q$, the GPD $E(x,0,0)$ requires only that $L_z^q$ is nonzero
on one side of the matrix element. This illustrates why we were only able to
derive a bound on the probability for nonzero $L_z^q$ that depends
quadratically 
on $E(x,0,0)$, which is why our bound is so weak. In contradistinction
$L_z^q$ obtained through the Ji relation would involve $E(x,0,0)$ linearly.
Nevertheless, we hope that the constraints derived in this work will
be of use in developing better models for light-cone wave functions 
describing these interesting observables.

{\bf Acknowledgments:}
We would like to thank M.~Diehl, M.~Oka and W.~Vogelsang for stimulating 
discussions. M.B. was partially supported by the DOE under grant number 
DE-FG03-95ER40965. M.B. would also like to thank T.-A.~Shibata for his
hospitality and support during part of this project. G.S. was partially
supported by the Japan Society for the Promotion of Science, the
Alexander-von-Humboldt Stiftung, and the Fonds voor Wetenschappelijk Onderzoek
-- Vlaanderen.

\end{document}